\documentclass[twocolumn,showpacs,preprintnumbers,amsmath,amssymb]{revtex4}

\usepackage{graphicx}
\usepackage{dcolumn}
\usepackage{bm}
\usepackage{epstopdf}

\begin{document}

\preprint{DRAFT: SUBMITTED TO AJP}

\title{Designing Effective Questions\\
for Classroom Response System Teaching}

\author{Ian D. Beatty}
\email{beatty@physics.umass.edu}
\homepage{http://umperg.physics.umass.edu}
\author{William J. Gerace}
\author{William J. Leonard}
\author{Robert J. Dufresne}
\affiliation{Scientific Reasoning Research Institute \& Department of Physics\\
University of Massachusetts\\
Amherst, MA 01003-9337 USA}

\date{\today}

\begin{abstract}
Classroom response systems (CRSs) can be potent tools for teaching physics. Their efficacy, however, depends strongly on the quality of the questions used. Creating effective questions is difficult, and differs from creating exam and homework problems. Every CRS question should have an explicit pedagogic purpose consisting of a content goal, a process goal, and a metacognitive goal. Questions can be engineered to fulfil their purpose through four complementary mechanisms: directing students' attention, stimulating specific cognitive processes, communicating information to instructor and students via CRS-tabulated answer counts, and facilitating the articulation and confrontation of ideas. We identify several tactics that help in the design of potent questions, and present four ``makeovers'' showing how these tactics can be used to convert traditional physics questions into more powerful CRS questions.
\end{abstract}


\maketitle

\section{Introduction}

Electronic classroom response systems (CRS) such as eInstruction {\em CPS}, InterWrite (formerly EduCue) {\em PRS}, and {\em H-ITT} can be powerful tools in the service of physics instruction.\cite{Dufresne:1996ct,Hake:1998ie,Mazur:1997pi,Penuel:2004cw,Roschelle:2004nc,Zollman:2005ec} They are merely tools, however, not a ``magic bullet.'' To significantly impact student learning, a CRS must be employed with skill in the service of sound, coherent pedagogy. This is not easy.

In a research project entitled {\em Assessing-to-Learn}, we helped high school teachers learn to teach physics with a CRS and studied their progress and difficulties.\cite{Dufresne:2000as,Dufresne:2004a2,Feldman:2003a2} We have also used a CRS in our own university physics teaching, helped others learn to do so, and designed and tested CRS questions for more than ten years.\cite{Dufresne:1996ct} Our experience spans a broad array of contexts: high school classes, introductory univerisity classes for non-science and science majors, upper-level university classes for physics majors, and workshops for in-service physics teachers and for science graduate students. We've used a CRS with classes ranging in size from fewer than twenty to over two hundred students. We've taught traditional physics material, and also ``conceptual'' physics for non-scientists, general science, and science pedagogy.

We can state unequivocally that learning to operate the technology is the easiest part of becoming facile with CRS-based instruction. More difficult challenges include creating and adapting suitable questions, cultivating productive classroom discourse, and integrating CRS use with the rest of the course, with curricular materials, and with external constraints.\cite{Feldman:2003a2}

Many who try teaching with a CRS discover that creating or finding ``good'' questions is more difficult than it at first appears. The characteristics of effective CRS questions are quite different from those of good exam questions, homework problems, and in-class worked examples. The vast archives of questions and problems that instructors accumulate over years of teaching, or find in standard textbooks, therefore offer little assistance to the new CRS user. Few collections specifically designed for CRS-based teaching exist.

One relatively widely-known collection of CRS questions is that contained in Eric Mazur's book {\em Peer Instruction}.\cite{Mazur:1997pi} Mazur has popularized CRS use in physics instruction, and the questions in his book make a useful starting point. However, we have found this collection to be insufficient for two reasons. One reason is that the questions have been designed to support Mazur's particular goals for CRS use, and are not optimal for the more ambitious approach we use and advocate. Mazur's {\em Peer Instruction} method consists of brief lecture-style presentations on key points of physics, each followed by a short conceptual question. Students are asked to formulate their own answers and then convince their peers of the correctness of their answer. Mazur argues that this process ``forces students to think though the arguments being developed, and\ldots provides them (as well as the teacher) with a way to assess their understanding of the concept'' (Ref. \onlinecite{Mazur:1997pi}, p. 10). He also suggests that knowledge ``diffuses'' among the students, helping to spread correct ideas.

While this way of using a CRS is viable and valuable, we find that even more dramatic improvements in teaching and learning can occur by inverting the paradigm. Rather than following mini-lectures with CRS-based quizzing, we use a CRS-powered ``question cycle'' as the core of in-class instruction, making question posing, pondering, answering, and discussing the vehicle of learning. Micro-lectures are injected only occasionally and when immediate circumstances warrant. Furthermore, we use a CRS to develop more than just conceptual understanding. We also target the development of cognitive skills, analysis and problem solving ability, and productive student metacognition about physics, learning, and thinking. Our approach, called {\em Question-Driven Instruction}, is summarized below in Subsection~\ref{sec:role} and described in more detail elsewhere.\cite{Beatty:2004ec,Beatty:2005ar,Dufresne:1996ct,Dufresne:2000as}

The other reason that the questions in {\em Peer Instruction} and similar collections are insufficient (although useful) is that instructors cannot use a question effectively if they do not appreciate its goals and design logic. They often fail to take advantage of the question's latent potential or unwittingly sabotage its effect. Instructors who would use a CRS well need a generalized understanding of what makes CRS questions succeed or fail, how to evaluate questions, and how to invent or adapt questions to meet their personal situation, objectives, and style. To put it bluntly, instructors should know how to create and modify, not just use, CRS questions. Although a few isolated question-design techniques have been developed and publicized,\cite{Li:2004ic} we are aware of no comprehensive or systematic framework for developing and evaluating CRS questions.

This paper addresses that need. Section~\ref{sec:theory} summarizes our vision of the central role that CRS use can play within physics instruction, and lays out a framework for thinking about and designing CRS questions. Section~\ref{sec:tactics} describes a selection of specific tactics that can be employed when designing questions. Section~\ref{sec:examples} presents four ``makeovers'': examples of traditional physics questions, together with improved variations that implement some of the tactics of Section~\ref{sec:tactics}. Section~\ref{sec:summary} summarizes the paper and offers some closing comments.

\section{\label{sec:theory}Theory: Goals and Mechanisms}

A general framework for thinking about question design must address the role questions will play within a course, the specific goals a question can be designed to attain, and the various mechanisms by which it can attain them.

\subsection{\label{sec:role}Role: What Part do Questions Play?}

We advocate a model of CRS-based teaching that we call {\em Question-Driven Instruction} (QDI). In this model, posing questions via CRS does more than augment traditional instruction: it forms the very core of the instructional dynamic. Our primary in-class goal is not to ``lecture'' or present information. Rather, we seek to help students explore, organize, integrate, and extend their knowledge. Students receive their primary exposure to new material from textbooks, multimedia, and other out-of-class resources.

\begin{figure}
\includegraphics{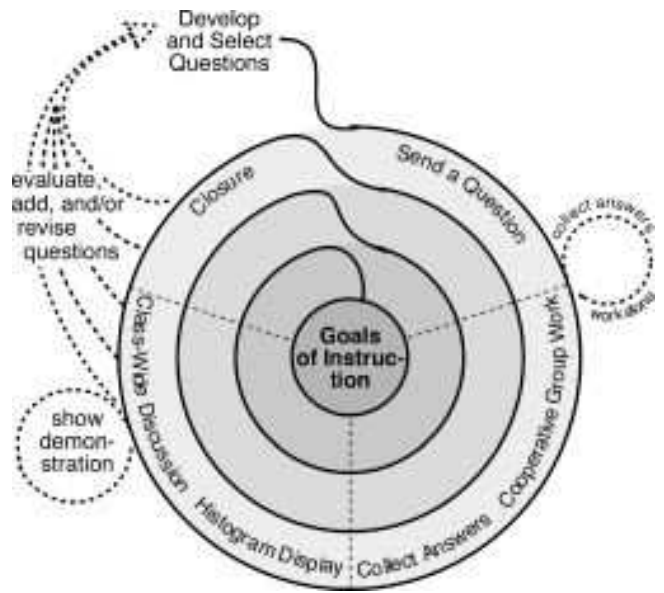}
\caption{\label{fig:qCycle} The {\em question cycle} used for Question-Driven Instruction with a classroom response system.}
\end{figure}

In-class activity is organized around a {\em question cycle} (Fig.~\ref{fig:qCycle}).\cite{Dufresne:2000as} We begin the cycle by presenting a question or problem to the class, generally without preamble, and allow a few minutes for students to discuss it in small groups. Typically, students within a group will argue their various opinions and intuitions, work out a solution if required, and continue discussing and elaborating until satisfied with their answer. Students then key in their responses. We view and display an instant histogram showing the class-wide distribution of responses. Without revealing which responses are superior, we then moderate a class-wide discussion, asking for volunteers to explain the reasoning behind each. With deft management, this process can be turned into a lively interchange of ideas and arguments between students. Based on the thinking students reveal during discussion, we can follow up with general observations, a brief micro-lecture, a related CRS question, or whatever else is necessary for closure; at this point, a few well-chosen comments can often precipitate significant learning. We typically repeat the entire cycle three or four times per 50-minute class.

Three aspects of the cycle are worth stressing here. First, questions are presented to students in a way that encourages significant cogitation, rather than just memory recall or execution of practiced skills. Second, questions are accompanied by extensive discussion: within small groups before answers are collected, and by the whole class afterward. Third, the instructor continually probes for and adjusts to the students' learning needs---a practice we call ``agile teaching.''

\subsection{Goals: What Should the Question Accomplish?}

We strongly believe that every CRS question used in class should serve an explicit pedagogic objective. By ``pedagogic objective,'' we mean more than just a particular piece of physics content. For maximum benefit, we maintain that every question should have a threefold purpose consisting of a content goal, a process goal, and a metacognitive goal.

\underline{\textbf{Content goal}:} A question's content goal is determined by answering the question, ``What piece(s) of subject material do we want to illuminate?'' This dimension of a question's purpose is the most obvious, and needs little discussion other than to suggest that concepts, principles, and their interrelationships make the best fodder for productive questions. Since QDI cannot explore all aspects of a subject in the classroom time allotted a typical course, we focus question cycle iterations on the foundational ideas, core principles, crucial distinctions, and conceptual organization of the material at hand. With a robust understanding of these, students are well prepared to learn advanced topics and special cases through reading and homework assignments.

\underline{\textbf{Process goal}:} A question's process goal is chosen by answering the question, ``What cognitive skills do we want students to exercise?'' If the content goal refers to {\em what} physics material students must use to answer the question, the process goal refers to {\em how} they must use it. One might also call a process goal a ``cognitive goal.''

In addition to knowledge about physics, expert physicists possess a wide range of skills that make their knowledge useful in various situations. We have identified twelve {\em habits of mind} that successful physicists practice and that students should be encouraged to develop.\cite{Dufresne:2000as} For convenience, we separate them into ``basic'' and ``advanced'' sets, summarized in Table~\ref{tab:HoM}. A question's process goal can be characterized by the habits of mind it exercises.

\begin{table}
\caption{\label{tab:HoM}``Habits of mind'' that expert physicists possess and students should develop.}
\begin{ruledtabular}
\begin{tabular}{ll}
Basic&Advanced\\
\hline
Seek alternative representations&Generate multiple solutions\\
Compare \& contrast&Categorize \& classify\\
Explain, describe \& depict&Discuss, summarize \& model\\
Predict \& observe&Strategize, justify \& plan\\
Extend the context&Reflect \& evaluate\\
Monitor \& refine communication&Think about thinking \& learning\\
\end{tabular}
\end{ruledtabular}
\end{table}

A crucial activity for students to engage in, spanning and integrating many habits of mind, is {\em analysis}: understanding a situation by identifying the essential concepts and their relationships, and reasoning with these to draw conclusions.\cite{Leonard:2001sp} The practice of analysis develops robust conceptual understanding and connects it to successful problem-solving ability. As part of their process goal, questions should frequently require students to perform qualitative analysis in pursuit of an answer.

\underline{\textbf{Metacognitive goal}:} A question's metacognitive goal is chosen by answering the question, ``What beliefs about learning and doing physics do we wish to reinforce?'' Everything that occurs within a course expresses, explicitly or implicitly, a perspective on physics, thinking, learning, teaching, and how the educational ``game'' should be played. To cite just a few of the many issues this includes: Is physics about memorizing and applying rules and equations, or about reasoning and making sense of the physical world? Should students study and work in isolation, or is learning a cooperative and social process? Is attention to detail important, or is getting the general idea sufficient? Should conscientious and able students expect to understand material the first time it is presented, or is confusion, resolution, and multi-pass learning necessary? How much self-directed activity should learning physics require? The answers to these questions may be obvious to most instructors, but they are not to students; consistency of message is crucial.

The more constructive students' metacognitive perspective is, the more efficiently they can learn what we are trying to teach. By influencing their perspective, we can significantly enhance learning in our courses, and help to prepare students for future learning throughout and beyond school. And recent thinking on the transfer of knowledge suggests that ``preparation for future learning'' is the most durable learning outcome our instruction is likely to achieve.\cite{Bransford:1999rt,Bransford:1999hp}

\subsection{Mechanisms: How Can a Question Accomplish its Goals?}

A QDI question can fulfil its pedagogic purpose by way of four different general mechanisms or ``channels'': through focusing students' attention by posing the question, through stimulating cognitive processes as students ponder the question, through feedback provided to students and instructor by collective viewing of the response histogram, and through articulation and confrontation of ideas during discussion.

Questions can have a very powerful effect on students merely by being posed and pondered. The first two mechanisms can be thought of as ``What are they thinking about?'' and ``How are they thinking about it?'' A question can direct students' attention to specific facts, features, ideas, conflicts, or relationships, bringing issues to conscious awareness. Sometimes merely looking at an idea, or looking at it from the right angle, is enough to spark understanding. Other times, it is a necessary but insufficient first step. Also, a question can stimulate students to exercise specific cognitive processes: habits of mind and the general practice of analysis. No question can force students to engage in any particular cognitive process, of course; mental engagement is always voluntary. However, the design of a question can necessitate certain processes to reach a successful answer, and can make the need for certain processes relatively obvious.

The third and most obvious mechanism by which a CRS question can serve pedagogic ends is by communicating information about student responses. By seeing the histogram of answers entered, the instructor learns about students' understanding, and students learn about their classmates' thinking. If this information is not merely noted, but actually used by the instructor or students to make subsequent teaching and learning decisions, then response system use constitutes {\em formative assessment}: assessment to enhance, rather than evaluate, learning. And formative assessment is perhaps the most effective instructional ``innovation'' ever studied.\cite{Bell:2001fa,Black:1988ib,Black:1988fa,Boston:2002fa,Hobson:1997fa}

Inter-student and student-instructor discussion is the fourth mechanism by which questions can fulfil their design objectives. The discussion that accompanies use of a question---within small groups before answering and class-wide after the histogram is displayed---is crucial to effective QDI. One reason is that the act of articulating beliefs, perceptions, assumptions, expectations, understanding, and reasoning is inherently valuable to students. Thinking is often ill-formed, nebulous, and inconsistent. When a student must cast such thinking into language, especially the precise language of physics, such deficiencies become evident and must be redressed.

Another reason is that discussion involves a confrontation of different perceptions, different analyses, and different conclusions. Exposing students to their classmates' thinking challenges their own and promotes learning. Arguing and reconciling differences promotes yet more. Telling students what to think is notoriously ineffective; eliciting their thinking, confronting it with alternatives, and seeking resolution works better.

Yet another reason is that whole-class discussion can reveal to the instructor far more about students' understanding and difficulties than any single histogram, no matter how informative the question's distracters. We use these discussions to actively inform ourselves of the nature and causes of our students' errors. This also is a kind of formative assessment.

\section{\label{sec:tactics}Tactics: Implementing the Theory}

The previous section presented a general framework for thinking about the design of questions for Question-Driven Instruction using a classroom response system. In this section, we present some specific tactics (listed in Table~\ref{tab:tactics}) that can be used to implement the framework. These are merely a helpful, representative set; others certainly exist. They have been grouped according to which of the four mechanisms of effect they employ. Our convention is to present tactic names {\em in italics} and habits of mind ``in quotes.'' (Some habits of mind are also tactic names; the formatting indicates which is meant in a particular context.) For many of the tactics, we have indicated which of the makeovers in Section~\ref{sec:examples} employ them.

\begin{table}
\caption{\label{tab:tactics}Question design tactics.}
\begin{ruledtabular}
\begin{tabular}{l}
{\em Tactics for directing attention and raising awareness:}\\
Remove inessentials\\
Compare \& contrast\\
Extend the context\\
Re-use familiar question situations\\
Oops-go-back\\
\hline
{\em Tactics for stimulating cognitive processes:}\\
interpret representations\\
Compare \& contrast\\
Extend the context\\
Identify a set or subset\\
Rank variants\\
Reveal a better way\\
Strategize only\\
Include extraneous information\\
Omit necessary information\\
\hline
{\em Tactics for formative use of response data:}\\
Answer choices reveal likely difficulties\\
Use ``none of the above''\\
\hline
{\em Tactics for promoting articulation discussion:}\\
Qualitative questions\\
Analysis \& reasoning questions\\
Multiple defensible answers\\
Require unstated assumptions\\
Trap unjustified assumptions\\
Deliberate ambiguity\\
Trolling for misconceptions\\
\end{tabular}
\end{ruledtabular}
\end{table}

\subsection{Tactics for Directing Attention and Raising Awareness}

\textbf{\textsl{Removing inessentials}} from a question is a general, obvious, and often neglected tactic for focusing students' attention where we want it. By this, we mean removing anything inessential to the instructor's pedagogic purpose for the question (not necessarily to the students' efforts to answer it). For example, we avoid having potentially distracting features in the question's situation or potentially distracting steps in the thinking students must do to reach an answer. A question with a quantitative point, for example, is only weakened by requiring quantitative calculations that may distract students and divert their cognitive resources. Makeover C in Section~\ref{sec:examples} demonstrates this tactic.

\textbf{\textsl{Compare and contrast}} is another tactic for directing attention and awareness. By having students compare two things, their attention will naturally be drawn to the differences between them. One way to implement this is to pose a question that has students comparing multiple situations---physical arrangements, processes, conditions, etc.---to categorize or order them. Another is to describe a situation and ask about the effect of changing some aspect of it. A third is to present a sequence of two or more CRS questions, in which the situations are identical but the query varies, or in which the same query is made about slightly different situations. Makeover A demonstrates a question comparing two situations.

\textbf{\textsl{Extending the context}} of a known idea or skill is a habit of mind, and also a valuable question design tactic. By asking a familiar question about an unfamiliar situation, students' attention is drawn to the ways in which the new situation differs from known ones and to the relevance of these differences, and broadens students' comprehension of the relevant ideas. Students expand their understanding of concepts beyond the limited scope they are initially encountered in by seeing them in varied contexts. For example, when students begin to grasp the procedure for finding the normal force of an object on a flat horizontal surface, tilt or curve the surface. When they assimilate that, add additional forces on the object or give it a nonzero acceleration. Makeover A incorporates this tactic by asking a linear dynamics question about a situation generally used for rotational dynamics.

\textbf{\textsl{Re-using familiar question scenarios}} also has its place. Reading, digesting, and interpreting a question statement requires nontrivial cognitive resources from students---resources that could be spent on understanding the points we want to make with the question. The tactic of removing inessentials helps reduce this ``cognitive load'' effect, as does building new questions from situations and systems students have already come to terms with. Use of the {\em extending the context} and {\em re-using familiar question scenarios} tactics should be balanced: in general, new ideas should be introduced in familiar question contexts, while somewhat familiar ideas should be explored and developed in novel contexts.

\textbf{\textsl{Oops-go-back}} is an awareness-raising tactic involving a sequence of two related questions. The first is a trap: a question designed to draw students into making a common error or overlooking a significant consideration. The instructor allows students to respond, and then moves to the second question without much discussion. The second causes students to realize their mistake on the first. When students are ``burned'' this way by a mistake and discover it on their own, they are far more likely to learn from it than if they are merely warned about it in advance or informed that they have committed it. A simple example, suitable early during coverage of kinematics, would be to ask students for the velocity of some object moving in the negative direction or in two dimensions, with positive scalar answer choices including the object's speed, and also ``None of the above.'' Many, insufficiently attuned to the distinction between speed and velocity, will erroneously select the speed. Then, a second question asks about the object's speed, causing many students to consider how this question differs from the previous and realize their error.

\begin{figure}
\includegraphics{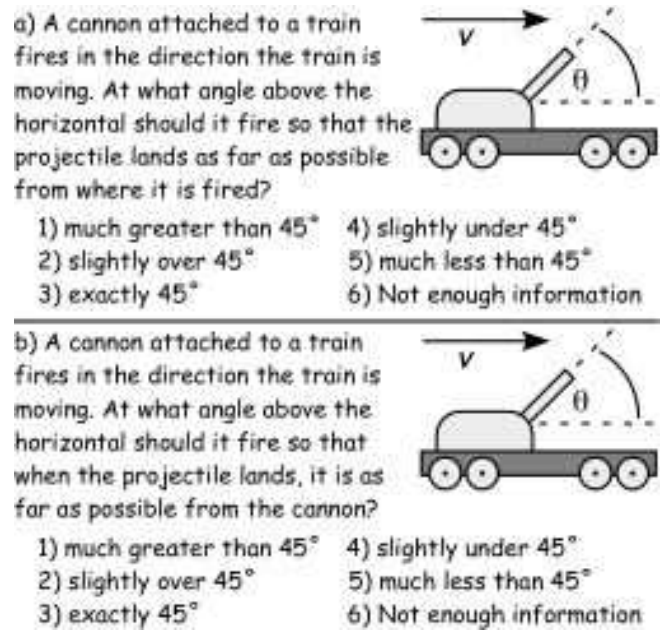}
\caption{\label{fig:oopsExample} A question pair exemplifying the {\em oops-go-back} design tactic.}
\end{figure}

This can be a subtle technique, so we will present a second, less trivial example. If we ask the question in Fig.~\ref{fig:oopsExample}a, many students will erroneously answer ``45 degrees.'' Without discussion, we then present the question in Fig.~\ref{fig:oopsExample}b. Many will realize that the answer to this one is ``45 degrees,'' be bothered by this, reconsider how the first question differed, and realize that they had neglected the cannon's velocity relative to the ground. While answering the first question, they had unwittingly answered the second.

\subsection{Tactics for Stimulating Cognitive Processes}

Specific question design tactics for stimulating cognitive processes are as varied as the spectrum of thinking skills they target. The fundamental rule is to ask questions which cannot be answered without exercising the desired habits of mind, and to avoid excess baggage that might distract students from the need to exercise them. Certain types of problems are helpful in this regard.

Many students are inordinately attached to algebraic representations of physics concepts, relationships, and situations and do not fully appreciate the utility of ``alternative'' representations such as graphs, free body and vector diagrams, and even verbal descriptions. Questions that require students to \textbf{\textsl{interpret representations}} are useful for remedying this and for developing the habit of mind ``seek alternative representations.'' The tactic is implemented by providing necessary information or answer options in such an alternative representation. For example, one might describe an object's motion with a graph and then ask a question about its behavior that requires students to recognize and interpret the information latent in aspects of the graph, such as a slope or the area under a curve. Alternatively, one might ask them to verbally describe the meaning of a mathematical equation, or to choose which of a set of vectors best describes some quantity. Makeover D does not require this of students, but it does rely on the tactic for the ``surprise'' solution revealed during discussion (cf. {\em reveal a better way}, below).

The \textbf{\textsl{compare and contrast}} and \textbf{\textsl{extend the context}} tactics for focusing students' attention, described above, are also useful for developing the habits of mind of those names. These are powerful question types with multiple benefits.

Some habits of mind are easy to target. ``Categorize and classify'' is promoted by presenting students with a set of situations, objects, processes, or other things and asking them to \textbf{\textsl{identify a set or subset}} meeting some criterion, or to \textbf{\textsl{rank variants}} according to some quality.

\textbf{\textsl{Constrain the solution}} is a tactic for exercising the habit of mind ``generate multiple solutions.'' This can be a ``positive constraint'' directing students to solve a problem via some particular approach (e.g., ``use the work-energy theorem''), or a ``negative constraint'' directing students not to use some particular approach (e.g., ``do it without solving for the acceleration''). Merely giving such a directive gets students to consider their activity from a strategic perspective.

\textbf{\textsl{Reveal a better way}} is another, less direct tactic for strengthening ``generate multiple solutions.'' One presents a question which students are likely to solve by a valid but difficult, tedious, error-prone, opaque, or otherwise non-optimal path. Then, during discussion, one can suggest a dramatically more elegant or simple solution. Makeover D is a classic example of this tactic.

\textbf{\textsl{Strategize only}} is a tactic for strengthening the habit of mind ``strategize, justify and plan.'' It is implemented by presenting a problem and asking students to identify the principle(s) and approach that would be most useful for reaching a solution, without actually solving the problem. This teaches students to think explicitly about problem solving and the set of strategies available to them. (The ``justify'' portion of that habit of mind is naturally developed during discussion of students' reasoning.) Makeover C demonstrates this.

\textbf{\textsl{Include extraneous information}} and \textbf{\textsl{omit necessary information}} are other tactics useful for developing ``strategize, justify and plan.'' They push students to consider explicitly what information is necessary to complete a strategy, rather than assuming every question provides exactly what is required and nothing more. (Note that {\em include extraneous information} is not inconsistent with {\em remove inessentials}, since deliberately extraneous information can be essential to the pedagogic purpose of the question. It is extraneous to students, but not to the instructor.)

Similar tactics can be imagined for other habits of mind. Once one has decided to target a specific cognitive facility and has escaped from the trap of always using standard calculate-an-answer questions, creating suitable question types is generally straightforward.

Cognitive processes are targeted not just by the intrinsic construction of the question, but also by classroom interaction surrounding its use. {\em Reveal a better way}, described above, relies on this. ``Monitor and refine communication'' is exercised any time communication within the classroom is explicitly discussed, perhaps after students have misinterpreted a question or a statement by the instructor. Similarly, ``think about thinking and learning'' ({\em metacognition}) is stimulated whenever the instructor asks students to explicitly consider their own thinking processes and learning behaviors.

\subsection{Tactics for Formative Use of Response Data}

As described above, the third general mechanism by which questions can fulfill their design objectives is by providing information to the instructor and students through the histogram of students' responses. To provide maximally useful information to the instructor, questions should be designed so that \textbf{\textsl{answer choices reveal likely student difficulties}}: 
common errors, misunderstandings, and alternative assumptions and interpretations. That way, by glancing at the histogram, we can quickly detect whether a particular one of these is prevalent in our class and decide whether to address it. In general, a response histogram is most useful to students and instructor when the spectrum of answers chosen is broad rather than narrowly peaked around one choice. (One exception is the first question of an {\em oops-go-back} pair, for which having a majority of students fall into the ``trap'' can be desirable.)

When interpreting students' responses, one must remember that any given answer can almost always be reached by more than one path or argument. Thus, having students explain their answers is vital. For this reason among others, we usually open the whole-class discussion for a question by proceeding systematically down the answer list, asking for (or cajoling) volunteers to present a rationale for each. We maintain a ``poker face'' throughout. After a student has argued for a particular answer, we ask if anyone has a different reason for the same answer.

We frequently \textbf{\textsl{include ``none of the above''}} (or ``not enough information'') as an answer choice, so as to learn about responses we might not have anticipated. We make this the ``correct'' or best answer often enough that students learn to take it seriously (often for {\it omit necessary information}, described above) and don't assume that they have made a mistake if they don't agree with one of the other options offered.

\subsection{Tactics for Promoting Articulation, Conflict and Productive Discussion}

We said earlier that the fourth mechanism of question efficacy---discussion---has students learn by articulating their thinking, confronting each others' thinking, and resolving differences. It also provides the instructor with valuable information about students' understanding, confusions, progress, and predilections. Not all questions lead to equally productive discussion. Questions that are most useful for this tend to be quite different from standard exam-type questions.

\textbf{\textsl{Qualitative questions}} are usually superior to quantitative ones for promoting articulation and argument. Quantitative questions lure students into thinking in terms of numbers, variables, and equations, which are difficult to communicate and discuss; qualitative questions promote discussion in terms of concepts, ideas, and general relationships. The final question versions in all four makeovers of Section~\ref{sec:examples} are qualitative.

\textbf{\textsl{Analysis and reasoning questions}}, requiring significant decision-making by students, similarly lead to better discussion and more valuable articulation than those requiring calculation or memory recall. (They also promote the development of analytic skills.) Makeovers B and C introduce elements of analysis, and D is a good analysis question (although it may initially seem like a straightforward algebra problem to students).

Questions with \textbf{\textsl{multiple defensible answers}} are useful for sowing dissension and generating productive discussion. Perhaps more than one answer is viable depending on how one chooses to interpret the question, or on what assumptions one consciously or unconsciously makes. Makeover B exemplifies this.

Similarly, one can design questions that \textbf{\textsl{require unstated assumptions}}, \textbf{\textsl{trap unjustified assumptions}}, or contain \textbf{\textsl{deliberate ambiguity}}. In addition to promoting disagreement and therefore profitable discussion, these have the benefit of sensitizing students to the multiple interpretations possible for many situations, to the importance of assumptions in physics reasoning, and to the criteria physicists use when evaluating assumptions. Makeover D introduces an unstated assumption (leading to {\em multiple defensible answers}).

\textbf{\textsl{Trolling for misconceptions}} is another useful tactic: engineering questions that deliberately catch students in likely misconceptions or undesirable ``alternative conceptions.'' Such questions further the content goal of helping students become aware of and escape the particular misconception, improving their physics knowledge. They further the metacognitive goal of putting students on the alert for misconceptions in general. They also tend to promote argument, sometimes impassioned. Makeovers A and B both target specific misconceptions.

How one conducts class discussion can be more important to the quality of the discussion than what questions are used. Emphasizing cogency of reasoning over correctness of results is crucial, as is stressing that the only ``bad'' answer is one that does not reflect a student's actual thinking. Other, more specific tactics for moderating discussion exist, but are outside the scope of this paper.

\section{\label{sec:examples}Examples}

Abstract advice, divorced from concrete examples, can be difficult to implement. In this section we present four ``makeovers'': case studies in which a traditional question is improved by incorporating some of the QDI question design tactics into it. Each is accompanied by a discussion of which tactics have been implemented.

\subsection{Newton's Second Law}

\begin{figure}
\includegraphics{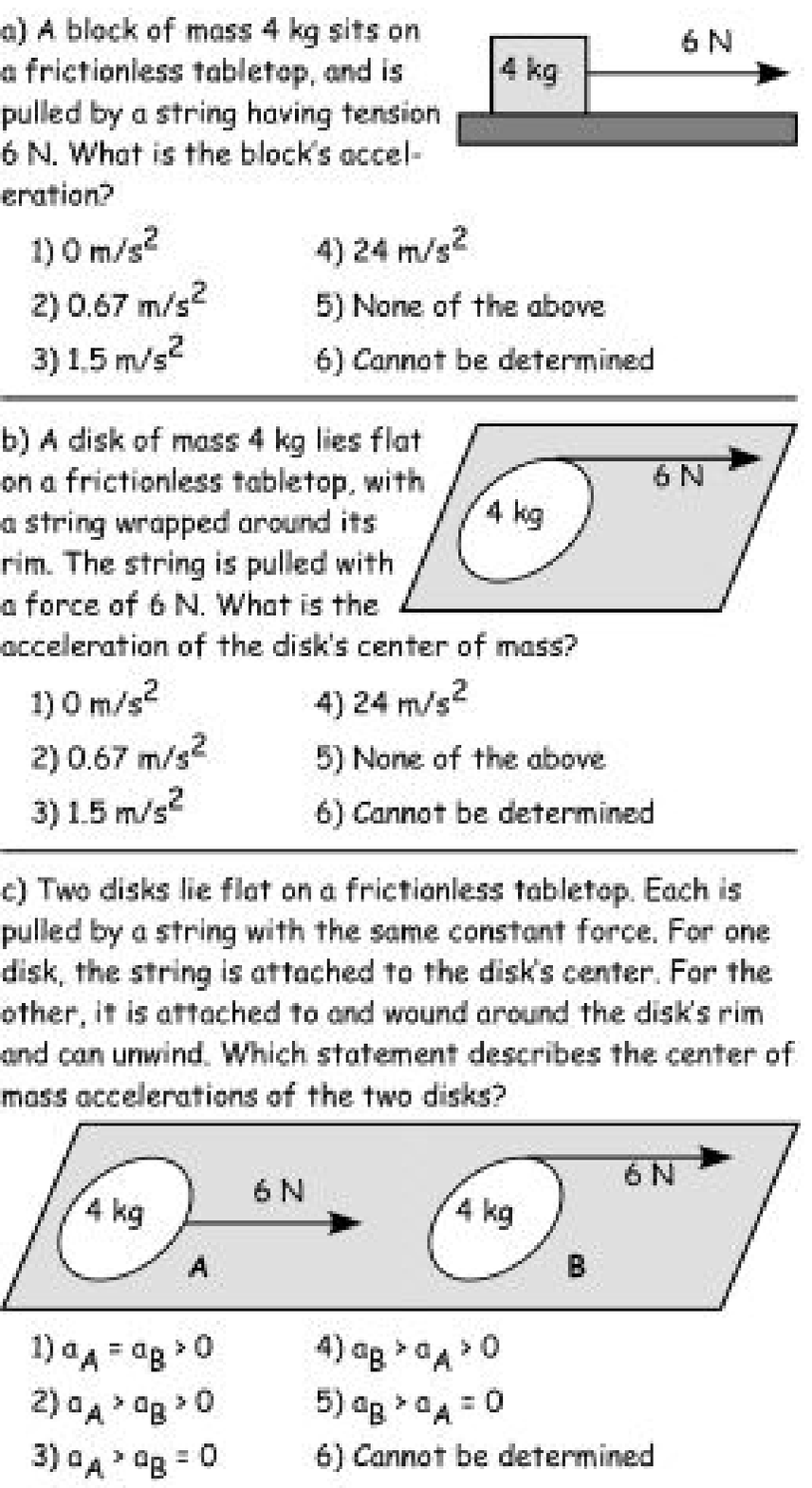}
\caption{\label{fig:example_A} Three variants of a question on Newton's Second Law in one dimension.}
\end{figure}

Fig.~\ref{fig:example_A}a shows a straightforward question on Newton's Second Law in one dimension. The variant in Fig.~\ref{fig:example_A}b requires the same content knowledge, but incorporates the tactic of {\em trolling for a misconception} to promote disagreement and argumentation. A common misconception among novice physics students is that $\tau = I\alpha$ somehow supersedes or modifies $F = ma$. By displaying the surface features of a rotational dynamics problem, this question will lure many students into the trap of thinking that because the disk rotates, some of the force is ``used up'' and the resulting linear acceleration will be less than Newton's Second Law alone would predict. Although requiring only Newton's Second Law in one linear dimension, the question would be appropriate for use after rotational dynamics has been introduced.

Note that the question wording does not explicitly ask for the {\em linear} acceleration of the disk. Most likely, some students will assume that the question is asking for angular acceleration and choose the ``cannot be determined'' answer. This allows the instructor to stress that ``acceleration,'' unqualified, means ``linear acceleration.'' If a large enough fraction of the class answers this way, we recommend clarifying the point, and then starting over without further discussion and allowing students to re-answer so that the question's primary intent can be realized. (To make the variant more effective at trapping this error, the disk's radius can be given, and some answer choices in appropriate units for angular acceleration can be provided.)

Fig.~\ref{fig:example_A}c shows a variant incorporating yet more question design tactics. {\em Compare and contrast} is used to focus students' attention on the effect of rotational motion on linear acceleration, and to practice the ``compare and contrast'' habit of mind. This variant is a {\em qualitative question} that trolls for the same misconception and interpretation error as the previous variant, but more effectively. It is a powerful tool for pushing students to articulate an intuitive misconception and to realize, wrestle with, and resolve the contradictions it leads to. (With deft handling, an instructor can use it to compare and contrast the ideas of force and acceleration with work and energy: although the two disks experience the same force and have the same acceleration, one gains more kinetic energy than the other in a given time interval.)

The variant would have been effective, and simpler, if the three answer choices were ``$a_{A} < a_{B}$,'' ``$a_{A} = a_{B}$,'' and  ``$a_{A} > a_{B}$.'' Both are implementations of {\em rank variants}. However, the set of distracters provided in variant b helps the instructor distinguish between students who think the disk will accelerate more slowly when rotating and those who think it will spin without translating at all, making use of {\em answer choices reveal likely difficulties}.

\subsection{Identifying Forces}

\begin{figure}
\includegraphics{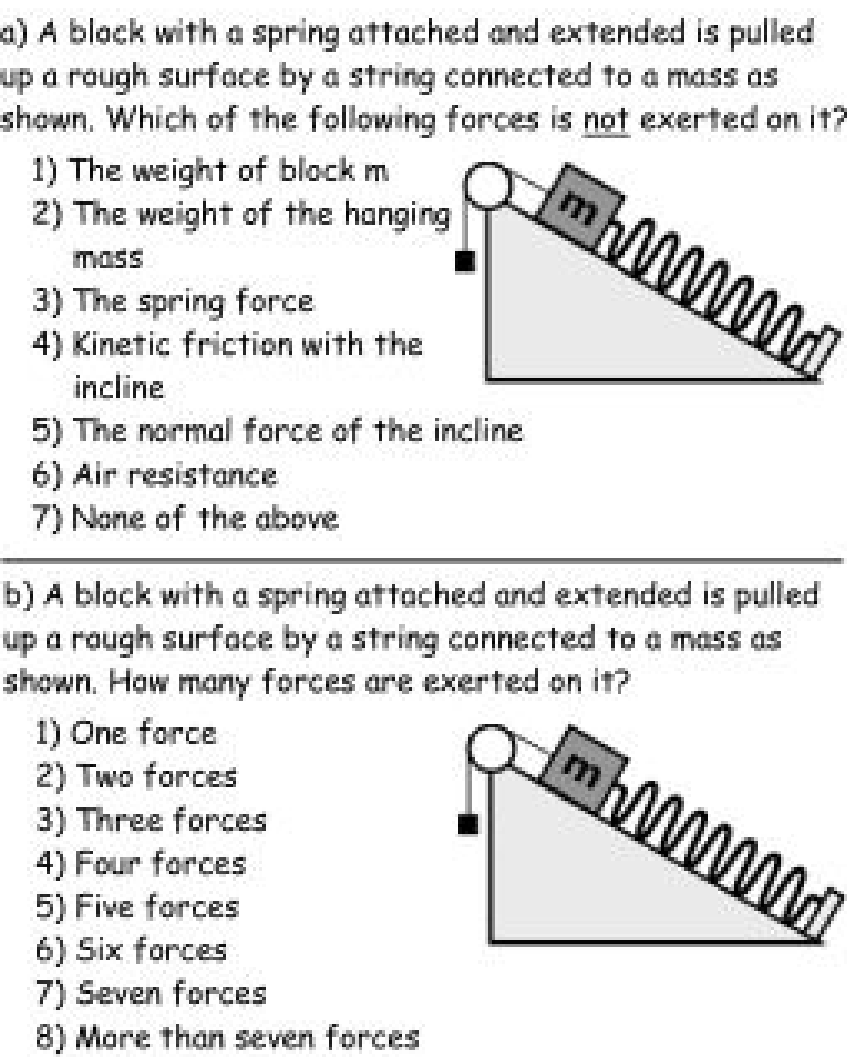}
\caption{\label{fig:example_B} Two variants of a question on identifying forces.}
\end{figure}

Fig.~\ref{fig:example_B}a shows a question targeting students' ability to identify the forces on a body. The question's content goal is to have students appreciate that apart from gravity (and other ``action at a distance'' forces not encountered during introductory mechanics), all forces acting on a body are caused by interactions with other bodies in direct contact with it. This is a {\em qualitative question} that {\em trolls for the misconception} that contact forces can be ``transmitted'' through an intermediate body to act between two separated bodies.

As written, the question is not bad, but it could be better. Consider the variant in Fig.~\ref{fig:example_B}b. It is still qualitative and trolls for the same misconception. However, it is open-ended and has students enumerate the forces (a modification of {\em identify a set or subset}) to unearth other, perhaps unanticipated misconceptions and errors. For example, one might discover that some students count the ``net force'' alongside other forces.

More importantly, it employs the {\em multiple defensible answers} tactic: choices 4 through 8 are all justifiable, depending on whether one treats the interaction between the plane and block as one contact force or as two (friction and normal), whether one neglects buoyancy and drag due to air, whether one includes ``silly'' but real forces like the gravitational effect of the moon, etc. (In fact, from a microscopic perspective, one can argue for an uncountably large and fluctuating number of forces due to molecular collisions.) This could also be considered an implementation of the {\em require unstated assumptions} or {\em deliberate ambiguity} tactics. Knowing what answer a student has picked conveys little information about their degree of understanding or about specific confusions they may have. Instead, the question serves to get students thinking about which forces are present without prompting them with specific forces. Then, during whole-class discussion, the instructor can ignore the answers chosen and instead discuss various possible forces in turn, arguing whether or not each merits counting.

This variant also addresses additional content goals. During discussion, the instructor can model a general procedure for identifying the forces on a body, and illuminate the various choices and conventions involved in identifying a ``force'' (for example, the convention of treating one contact force between two surfaces as two distinct, orthogonal forces: the ``normal'' and ``friction'' forces). In addition, the question makes an excellent platform for discussing the role of assumptions and approximations in physics and helping students learn when to include or neglect various forces.

In our minds, perhaps the most powerful aspect of this variant is its effectiveness at achieving the metacognitive goal of communicating to students that they should be concerned with reasoning, learning, and the cogency of their answers, and not with the correctness or incorrectness of any particular answer choice. Most students are deeply attached to the notion that every question has a ``right'' or ``best'' answer. We have found that the only way to make students abandon this and really focus on reasoning is to use questions like \ref{fig:example_B}b, for which it is patently obvious that several answers are defensible and can be ``correct'' or ``incorrect'' depending on the argument behind them.

\subsection{Energy and Angular Motion}

\begin{figure}
\includegraphics{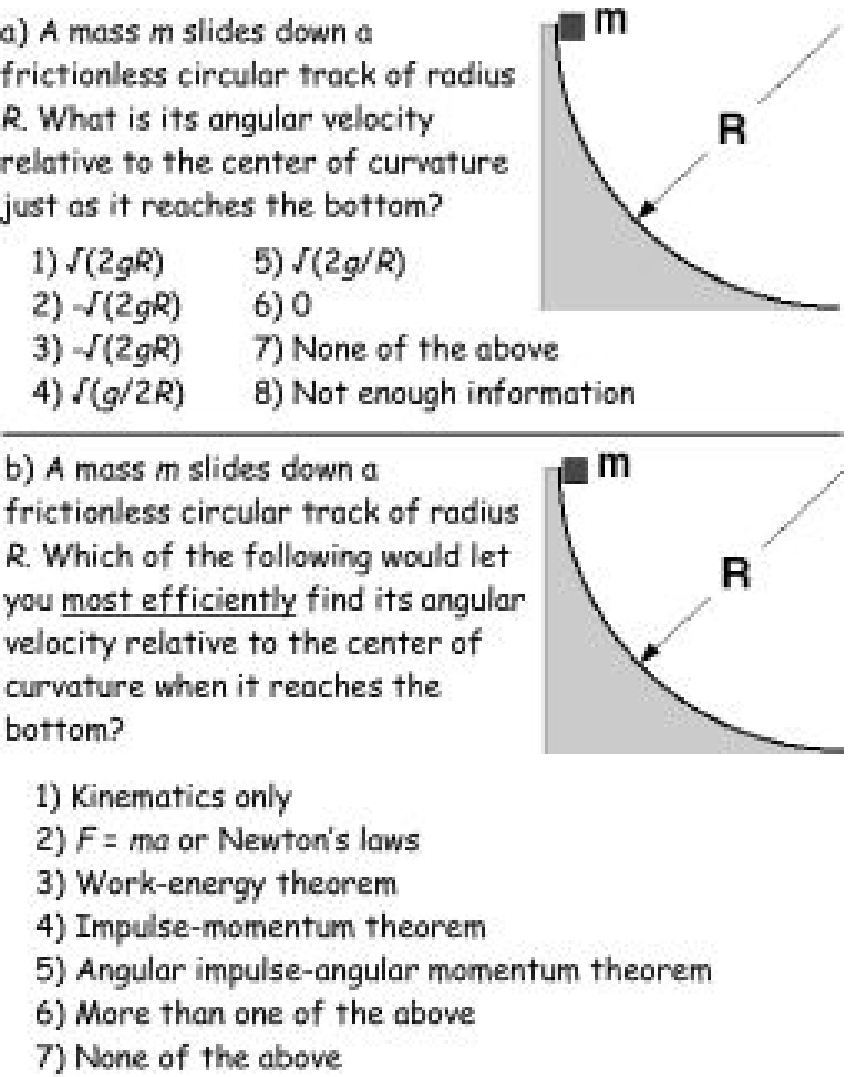}
\caption{\label{fig:example_C} Two variants of a question on strategic choices in problem solving.}
\end{figure}

Fig.~\ref{fig:example_C}a shows a question that helps students integrate their knowledge by requiring a mix of energy and angular motion ideas. To answer the question correctly, students must recognize the need for conservation of energy, apply it, and relate linear to angular motion.

The variant in Fig.~\ref{fig:example_C}b improves on the original by incorporating several question design tactics. Since the question's primary content goal is to develop students' ability to recognize the need for two different strategic steps using physics from two distinct topic areas (energy conservation and relating linear to angular motion), this variant is a {\em qualitative question} that {\em removes inessentials} to focus students' attention more effectively; students rarely pay sufficient attention to high-level, strategic aspects of problem solving when embroiled in equation manipulation. It uses the {\em strategize only} tactic. The question's phrase ``most efficiently'' can be considered a {\em deliberate ambiguity}: is efficiency defined in terms of number of lines of calculation required, number of principles involved, intricacy of thought entailed, or something else? Does efficiency depend on the skills of the learner? (If the question initiates a class-wide discussion of ``efficiency'' in problem solving, so much the better.)

\subsection{Kinematics}

\begin{figure}
\includegraphics{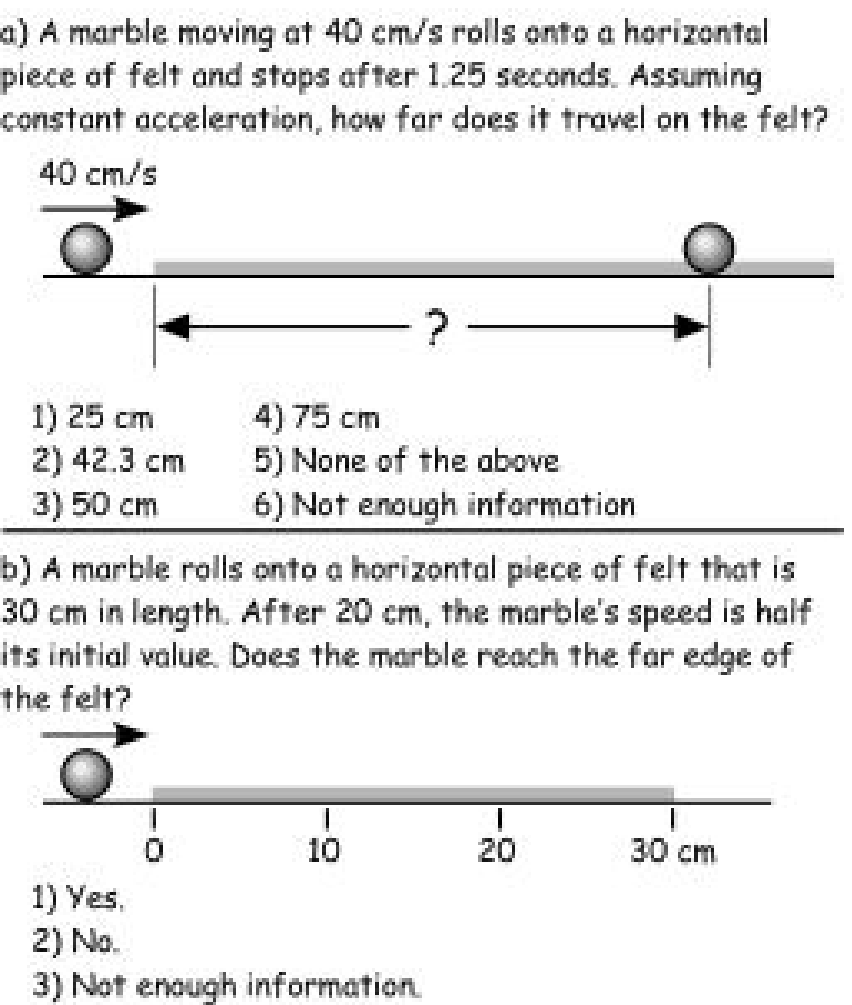}
\caption{\label{fig:example_D} Two variants of a kinematics question.}
\end{figure}

Fig.~\ref{fig:example_D}a is a relatively straightforward kinematics question. It is nontrivial, in that students aren't given the acceleration and must determine it first before calculating the distance traveled. Alternatively, students can determine the average velocity and multiply that by the time. The question requires some strategic decision-making.

The variant in Fig.~\ref{fig:example_D}b is similar in content. If approached algebraically, it is also similar in difficulty. However, this variant is a {\em qualitative question}, making it more amenable to analysis and reasoning and more suitable for discussion. By omitting the statement ``Assuming a constant acceleration,'' we have employed the {\em require unstated assumptions} tactic (and {\em multiple defensible answers} as a result) and opened up the possibility of discussing whether and how the constancy of the acceleration matters.

More interestingly, this variant permits the instructor to {\em reveal a better way} of answering the question. Assume constant acceleration and sketch a graph of velocity vs. time. By identifying the area under the line (the time integral) with the distance traveled, one can use simple geometry to see that when the marble has slowed to half its original velocity, it has traveled three-quarters of the distance it will cover before coming to rest. If it reaches half speed after 20 cm/s, it must stop at 4/3 that distance, which is before the end of the felt. This alternate solution involves {\em interpreting representations}, and the variant is a better {\em analysis and reasoning question} than variant a.

\section{\label{sec:summary}Summary \& Remarks}

In the Question Driven Instruction approach, using a classroom response system to pose, collect answers for, and discuss questions forms the core of in-class instruction. The success of the approach depends in part on the quality of the questions used. Effective questions should be designed with an explicit, threefold pedagogic objective consisting of a content goal, a process (cognitive) goal, and a metacognitive goal. The content goal is the topic material to be addressed; it should generally be conceptual and foundational in nature, and should frequently integrate ideas from various portions of the curriculum. The process goal is the set of cognitive skills to be developed, and can be thought of in terms of twelve ``habits of mind'' and the general practice of qualitative analysis. The metacognitive goal is the perspective about learning or physics to be reinforced.

A question can fulfil its threefold pedagogic objective through four complementary mechanisms. The posing of the question can focus students' attention on particular issues. Students' pondering of the question can stimulate particular cognitive skills. Displaying the answer histogram can convey information about student knowledge and thinking to classmates and to the instructor. And discussion, both small-group and whole-class, can impact students and inform the teacher as students struggle to articulate and defend their thinking and confront others' perceptions, interpretations, assumptions, and reasoning.

Questions can be deliberately engineered for maximal learning, and we have identified some tactics such as {\em remove inessentials}, {\em compare and contrast}, {\em interpret representations}, and {\em strategize only} that can be used in the design of powerful questions. Often a question in the ``standard'' style can be improved through minor modifications that take advantage of one or more of these tactics.

Even with an explicit framework such as the one presented here, designing effective questions is challenging and time-consuming, and---like any other nontrivial skill---requires practice. A repository of well-designed questions can be quite helpful, and we have made many of our questions available through both a public website\cite{Beatty:2000al} and an annotated commercial product.\cite{Beatty:2005al} However, to teach most effectively with another person's questions, one should understand the goals and design logic of each one.\cite{Feldman:2003a2} The framework herein helps to analyze existing questions as well.

We reiterate that well designed questions are merely a tool, one component of the QDI approach. Pedagogy---how the instructor uses questions to interact with students in the classroom---is more important.\cite{Beatty:2004ec,Beatty:2005ar,Dufresne:2000as} Nevertheless, lack of effective questions can be a serious and frustrating barrier to teachers seeking to learn and practice QDI. We believe the framework and question design tactics presented here can help overcome this barrier.

Although a classroom response system is a tremendously useful tool for implementing QDI, it is not essential to the philosophy underlying QDI. Many of the ideas presented here can be productively employed without technology, especially in small, highly interactive classes.

QDI is a type of formative assessment. By is very nature, formative assessment tends to be self-correcting: the feedback it provides to the practitioner about student learning can, if studied attentively, reveal implementation weaknesses and improve practice over time. Therefore, our most important piece of advice regarding QDI is to {\em pay critical attention to what happens when you do it}. Your students are your best teachers.

\begin{acknowledgments}
Ideas and perspectives presented in this paper have been developed during research projects supported by U.S. National Science Foundation grants DUE-9453881, ESI-9730438, and ROLE-0106771.
\end{acknowledgments}


\begin{thebibliography}{22}
\expandafter\ifx\csname natexlab\endcsname\relax\def\natexlab#1{#1}\fi
\expandafter\ifx\csname bibnamefont\endcsname\relax
  \def\bibnamefont#1{#1}\fi
\expandafter\ifx\csname bibfnamefont\endcsname\relax
  \def\bibfnamefont#1{#1}\fi
\expandafter\ifx\csname citenamefont\endcsname\relax
  \def\citenamefont#1{#1}\fi
\expandafter\ifx\csname url\endcsname\relax
  \def\url#1{\texttt{#1}}\fi
\expandafter\ifx\csname urlprefix\endcsname\relax\def\urlprefix{URL }\fi
\providecommand{\bibinfo}[2]{#2}
\providecommand{\eprint}[2][]{\url{#2}}

\bibitem[{\citenamefont{Dufresne et~al.}(1996)\citenamefont{Dufresne, Gerace,
  Leonard, Mestre, and Wenk}}]{Dufresne:1996ct}
\bibinfo{author}{\bibfnamefont{R.~J.} \bibnamefont{Dufresne}},
  \bibinfo{author}{\bibfnamefont{W.~J.} \bibnamefont{Gerace}},
  \bibinfo{author}{\bibfnamefont{W.~J.} \bibnamefont{Leonard}},
  \bibinfo{author}{\bibfnamefont{J.~P.} \bibnamefont{Mestre}},
  \bibnamefont{and} \bibinfo{author}{\bibfnamefont{L.}~\bibnamefont{Wenk}},
  \bibinfo{title}{``Classtalk: A classroom communication system for active learning,''}
  \bibinfo{journal}{J. Comput. High. Educ.}
  \textbf{\bibinfo{volume}{7}}, \bibinfo{pages}{3} (\bibinfo{year}{1996}).

\bibitem[{\citenamefont{Hake}(1998)}]{Hake:1998ie}
\bibinfo{author}{\bibfnamefont{R.}~\bibnamefont{Hake}},
  \bibinfo{title}{``Interactive-engagement vs. traditional methods: A six-thousand-student survey of mechanics test data for introductory physics courses,''}
  \bibinfo{journal}{Am. J. Phys.} \textbf{\bibinfo{volume}{66}},
  \bibinfo{pages}{64} (\bibinfo{year}{1998}).

\bibitem[{\citenamefont{Mazur}(1997)}]{Mazur:1997pi}
\bibinfo{author}{\bibfnamefont{E.}~\bibnamefont{Mazur}},
  \emph{\bibinfo{title}{Peer Instruction: A User's Manual}}
  (\bibinfo{publisher}{Prentice Hall}, \bibinfo{address}{Upper Saddle River,
  NJ}, \bibinfo{year}{1997}).

\bibitem[{\citenamefont{Penuel et~al.}(2004)\citenamefont{Penuel, Roschelle,
  Crawford, Shechtman, and Abrahamson}}]{Penuel:2004cw}
\bibinfo{author}{\bibfnamefont{W.~R.} \bibnamefont{Penuel}},
  \bibinfo{author}{\bibfnamefont{J.}~\bibnamefont{Roschelle}},
  \bibinfo{author}{\bibfnamefont{V.}~\bibnamefont{Crawford}},
  \bibinfo{author}{\bibfnamefont{N.}~\bibnamefont{Shechtman}},
  \bibnamefont{and} \bibinfo{author}{\bibfnamefont{L.}~\bibnamefont{Abrahamson}}, 
  \bibinfo{title}{``CATAALYST workshop report: Advancing research on the transformative potential of interactive pedagogies and classroom networks,''}
  \bibinfo{type}{Workshop Report}
  \bibinfo{number}{P14566}, \bibinfo{institution}{SRI International}
  (\bibinfo{year}{2004}).

\bibitem[{\citenamefont{Roschelle et~al.}(2004)\citenamefont{Roschelle, Penuel,
  and Abrahamson}}]{Roschelle:2004nc}
\bibinfo{author}{\bibfnamefont{J.}~\bibnamefont{Roschelle}},
  \bibinfo{author}{\bibfnamefont{W.~R.} \bibnamefont{Penuel}},
  \bibnamefont{and} \bibinfo{author}{\bibfnamefont{L.}~\bibnamefont{Abrahamson}}, 
  \bibinfo{title}{``The networked classroom,''}
  \bibinfo{journal}{Educational Leadership}
  \textbf{\bibinfo{volume}{61}}, \bibinfo{pages}{50} (\bibinfo{year}{2004}).

\bibitem[{\citenamefont{Zollman and Rebello}(2005)}]{Zollman:2005ec}
\bibinfo{author}{\bibfnamefont{D.}~\bibnamefont{Zollman}} \bibnamefont{and}
  \bibinfo{author}{\bibfnamefont{N.~S.} \bibnamefont{Rebello}},
  \bibinfo{title}{``The Evolving Classroom Response System at KSU: Classtalk, PRS, PDAs,''}
  \emph{\bibinfo{booktitle}{130th National Meeting of the American Association
  of Physics Teachers}} (\bibinfo{address}{Albuquerque, NM}, \bibinfo{year}{2005}).

\bibitem[{\citenamefont{Dufresne and Gerace}(2004)}]{Dufresne:2004a2}
\bibinfo{author}{\bibfnamefont{R.~J.} \bibnamefont{Dufresne}} \bibnamefont{and}
  \bibinfo{author}{\bibfnamefont{W.~J.} \bibnamefont{Gerace}},
  \bibinfo{title}{``Assessing-to-Learn: Formative assessment in physics instruction,''}
  \bibinfo{journal}{Phys. Teach.} \textbf{\bibinfo{volume}{42}},
  \bibinfo{pages}{109} (\bibinfo{year}{2004}).

\bibitem[{\citenamefont{Dufresne et~al.}(2000)\citenamefont{Dufresne, Gerace,
  Mestre, and Leonard}}]{Dufresne:2000as}
\bibinfo{author}{\bibfnamefont{R.~J.} \bibnamefont{Dufresne}},
  \bibinfo{author}{\bibfnamefont{W.~J.} \bibnamefont{Gerace}},
  \bibinfo{author}{\bibfnamefont{J.~P.} \bibnamefont{Mestre}},
  \bibnamefont{and} \bibinfo{author}{\bibfnamefont{W.~J.}~\bibnamefont{Leonard}}, 
  \bibinfo{title}{``ASK-IT/A2L: Assessing student knowledge with instructional technology,''}
  \bibinfo{type}{Technical Report} \bibinfo{number}{UMPERG-2000-09},
  \bibinfo{institution}{University of Massachusetts Physics Education Research Group}
  (\bibinfo{year}{2000}).

\bibitem[{\citenamefont{Feldman and Capobianco}(2003)}]{Feldman:2003a2}
\bibinfo{author}{\bibfnamefont{A.}~\bibnamefont{Feldman}} \bibnamefont{and}
  \bibinfo{author}{\bibfnamefont{B.}~\bibnamefont{Capobianco}},
  \bibinfo{title}{``Real-time formative assessment: A study of teachers' use of an electronic response system to facilitate serious discussion about physics concepts,''}
  \emph{\bibinfo{booktitle}{Annual Meeting of the American Educational Research
  Association}} (\bibinfo{address}{Chicago, IL}, \bibinfo{year}{2003}).

\bibitem[{\citenamefont{Beatty}(2004)}]{Beatty:2004ec}
\bibinfo{author}{\bibfnamefont{I.~D.} \bibnamefont{Beatty}},
  \bibinfo{title}{``Transforming student learning with classroom communication systems,''}
  \bibinfo{type}{Research Bulletin} \bibinfo{number}{ERB0403},
  \bibinfo{institution}{Educause Center for Applied Research}
  (\bibinfo{year}{2004}).

\bibitem[{\citenamefont{Beatty et~al.}(in press)\citenamefont{Beatty, Leonard,
  Gerace, and Dufresne}}]{Beatty:2005ar}
\bibinfo{author}{\bibfnamefont{I.~D.} \bibnamefont{Beatty}},
  \bibinfo{author}{\bibfnamefont{W.~J.} \bibnamefont{Leonard}},
  \bibinfo{author}{\bibfnamefont{W.~J.} \bibnamefont{Gerace}},
  \bibnamefont{and} \bibinfo{author}{\bibfnamefont{R.~J.}
  \bibnamefont{Dufresne}},
  \bibinfo{title}{``Question Driven Instruction: Teaching science (well) with an audience response system,''}
   in \emph{\bibinfo{booktitle}{Audience Response
  Systems in Higher Education: Applications and Cases}}, edited by
  \bibinfo{editor}{\bibfnamefont{D.~A.} \bibnamefont{Banks}}
  (\bibinfo{publisher}{Idea Group Inc.}, \bibinfo{address}{Hershey, PA}, \bibinfo{year}{in press}).

\bibitem[{\citenamefont{Li et~al.}(2004)\citenamefont{Li, Reay, and
  Bao}}]{Li:2004ic}
\bibinfo{author}{\bibfnamefont{P.}~\bibnamefont{Li}},
  \bibinfo{author}{\bibfnamefont{N.~W.} \bibnamefont{Reay}}, \bibnamefont{and}
  \bibinfo{author}{\bibfnamefont{L.}~\bibnamefont{Bao}},
  \bibinfo{title}{``Effects of in-class polling on student performance in learning physics,''}
  \emph{\bibinfo{booktitle}{129th National Meeting of the American Association
  of Physics Teachers}} (\bibinfo{address}{Sacramento, CA}, \bibinfo{year}{2004}).

\bibitem[{\citenamefont{Leonard et~al.}(2001)\citenamefont{Leonard, Gerace, and
  Dufresne}}]{Leonard:2001sp}
\bibinfo{author}{\bibfnamefont{W.~J.} \bibnamefont{Leonard}},
  \bibinfo{author}{\bibfnamefont{W.~J.} \bibnamefont{Gerace}},
  \bibnamefont{and} \bibinfo{author}{\bibfnamefont{R.~J.} \bibnamefont{Dufresne}}, 
  \bibinfo{title}{``Analysis-Based Problem Solving: Making analysis and reasoning the focus of physics instruction,''}
  \bibinfo{type}{Technical report} \bibinfo{number}{UMPERG-2001-12},
  \bibinfo{institution}{University of Massachusetts Physics Education Research Group}
  (\bibinfo{year}{2001}).
  \bibinfo{note}{Published in Spanish as ``Resoluti{\'o}n de
  Problemas Basada en el An{\'a}lisis: Hacer del an{\'a}lisis y del
  razonamiento el foco de la ense{\~n}anza de la f{\'i}sica,''
  Ense{\~n}anza de las Ciencias {\bf 20} (3, November): 387--400 (2002).}

\bibitem[{\citenamefont{Bransford and Schwartz}(1999)}]{Bransford:1999rt}
\bibinfo{author}{\bibfnamefont{J.~D.} \bibnamefont{Bransford}}
  \bibnamefont{and} \bibinfo{author}{\bibfnamefont{D.}~\bibnamefont{Schwartz}},
  \bibinfo{title}{``Rethinking transfer: A simple proposal with multiple implications,''}
  in \emph{\bibinfo{booktitle}{Review of Research in Education}}, edited by
  \bibinfo{editor}{\bibfnamefont{A.}~\bibnamefont{Iran-Nejad}}
  \bibnamefont{and} \bibinfo{editor}{\bibfnamefont{P.~D.}
  \bibnamefont{Pearson}} (\bibinfo{publisher}{American Educational Research
  Association}, \bibinfo{address}{Washington, D.C.}, \bibinfo{year}{1999}),
  vol.~\bibinfo{volume}{24}, pp. \bibinfo{pages}{61--100}.

\bibitem[{\citenamefont{Bransford et~al.}(1999)\citenamefont{Bransford, Brown,
  and Cocking}}]{Bransford:1999hp}
\bibinfo{author}{\bibfnamefont{J.~D.} \bibnamefont{Bransford}},
  \bibinfo{author}{\bibfnamefont{A.~L.} \bibnamefont{Brown}}, \bibnamefont{and}
  \bibinfo{author}{\bibfnamefont{R.~R.} \bibnamefont{Cocking}},
  \emph{\bibinfo{title}{How People Learn: Brain, Mind, Experience, and School}}
  (\bibinfo{publisher}{National Academy Press}, \bibinfo{address}{Washington,
  DC}, \bibinfo{year}{1999}).

\bibitem[{\citenamefont{Bell and Cowie}(2001)}]{Bell:2001fa}
\bibinfo{author}{\bibfnamefont{B.}~\bibnamefont{Bell}} \bibnamefont{and}
  \bibinfo{author}{\bibfnamefont{B.}~\bibnamefont{Cowie}},
  \bibinfo{title}{``The characteristics of formative assessment in science education,''}
  \bibinfo{journal}{Sci. Educ.} \textbf{\bibinfo{volume}{85}},
  \bibinfo{pages}{536} (\bibinfo{year}{2001}).

\bibitem[{\citenamefont{Black and Wiliam}(1988{\natexlab{a}})}]{Black:1988fa}
\bibinfo{author}{\bibfnamefont{P.}~\bibnamefont{Black}} \bibnamefont{and}
  \bibinfo{author}{\bibfnamefont{D.}~\bibnamefont{Wiliam}},
  \bibinfo{title}{``Assessment and classroom learning,''}
  \bibinfo{journal}{Assessment in Education: Principles, Policy and Practice}
  \textbf{\bibinfo{volume}{5}}, \bibinfo{pages}{7}
  (\bibinfo{year}{1988}{\natexlab{a}}).

\bibitem[{\citenamefont{Black and Wiliam}(1988{\natexlab{b}})}]{Black:1988ib}
\bibinfo{author}{\bibfnamefont{P.}~\bibnamefont{Black}} \bibnamefont{and}
  \bibinfo{author}{\bibfnamefont{D.}~\bibnamefont{Wiliam}},
  \bibinfo{title}{``Inside the black box: Raising standards through classroom assessment,''}
  \bibinfo{journal}{Phi Delta Kappan} \textbf{\bibinfo{volume}{80}},
  \bibinfo{pages}{139} (\bibinfo{year}{1988}{\natexlab{b}}).

\bibitem[{\citenamefont{Boston}(2002)}]{Boston:2002fa}
\bibinfo{author}{\bibfnamefont{C.}~\bibnamefont{Boston}},
  \bibinfo{title}{``The concept of formative assessment,''}
  \bibinfo{type}{Technical Report} \bibinfo{number}{ED470206},
  \bibinfo{institution}{ERIC Clearninghouse on Assessment and Evaluation} (\bibinfo{year}{2002}).

\bibitem[{\citenamefont{Hobson}(1997)}]{Hobson:1997fa}
\bibinfo{author}{\bibfnamefont{E.~H.} \bibnamefont{Hobson}},
  \bibinfo{title}{``Formative assessment: An annotated bibliography,''}
  \bibinfo{journal}{Clearing House} \textbf{\bibinfo{volume}{71}},
  \bibinfo{pages}{123} (\bibinfo{year}{1997}).

\bibitem[{\citenamefont{Beatty}(2000)}]{Beatty:2000al}
\bibinfo{author}{\bibfnamefont{I.~D.} \bibnamefont{Beatty}},
  \emph{\bibinfo{title}{Assessing-to-Learn project website}}
  (\bibinfo{year}{2000}), \url{<http://A2L.physics.umass.edu/>}.

\bibitem[{\citenamefont{Beatty et~al.}(2005)\citenamefont{Beatty, Leonard, and
  Gerace}}]{Beatty:2005al}
\bibinfo{author}{\bibfnamefont{I.~D.} \bibnamefont{Beatty}},
  \bibinfo{author}{\bibfnamefont{W.~J.} \bibnamefont{Leonard}},
  \bibnamefont{and} \bibinfo{author}{\bibfnamefont{W.~J.}
  \bibnamefont{Gerace}}, \emph{\bibinfo{title}{Assessing-to-Learn in the
  Classroom}} (\bibinfo{publisher}{Thomson Learning}, \bibinfo{year}{2005}),
  \url{<http://physics.brookscole.com/a2lc>}.

\end{thebibliography}


\end{document}